\documentclass[9pt,twocolumn,twoside]{opticajnl}
\journal{opticajournal}
\usepackage{siunitx}

\usepackage[section]{placeins}
\usepackage{subcaption}
\usepackage{dblfloatfix}
\usepackage{amsmath}
\usepackage{hyperref}
\usepackage[section]{placeins}

\usepackage{array}
\newcolumntype{C}[1]{>{\centering\arraybackslash}p{#1}}

\setboolean{shortarticle}{false}

\usepackage{lineno}

\title{Large field-of-view, distortion-corrected off-axis parabolic mirror relay microscope}

\author[1,*]{Meimei Liu}
\author[1,$\dagger$]{Yuqi Zhu}
\author[1,2]{Giorgio Gratta}

\affil[1]{Department of Physics, Stanford University, Stanford, CA 94305, USA}
\affil[2]{Hansen Experimental Physics Lab, Stanford University, Stanford, CA 94305, USA}
\affil[$\dagger$]{Now at Apple Inc.}

\affil[*]{hechen@stanford.edu}

\begin{abstract}  
Off-axis parabolic mirrors (OAPs) are occasionally desirable for specialized applications, but are known to introduce field-dependent astigmatic aberrations. In an experiment where optical tweezers are formed by OAPs, another OAP is added to form a relay configuration with an optional microscope, resulting in near diffraction-limited performance, with a resolution of $\qty{2.19}{\micro\meter}$. The severe radially non-uniform distortion incurred by this configuration, with a long path length and a large field of view, requires software corrections for the microscopic image. To avoid overfitting given the limited features available for calibration at the tweezers focal plane, a distortion model with a reduced set of parameters is selected based on simulation data. Applying the model to experimental data, an average residual error of $\qty{3.60}{\micro\meter}$ ($\qty{1.33}{\micro\meter}$) is achieved in object space after the simple relay (adding a $5 \times$ microscope) over a field of $\sim \qty{1}{\milli \meter} \times \qty{1}{\milli \meter}$ ($\sim \qty{200}{\micro \meter} \times \qty{200}{\micro \meter}$). The residual errors are likely dominated by diffraction artifacts in the features used for correction.

\end{abstract}

\setboolean{displaycopyright}{false}

\begin{document}

\maketitle

\section{Introduction}

Off-axis parabolic mirrors (OAPs) are desirable elements that provide a diffraction-limited focus from a collimated beam, while avoiding beam obscuration and chromatic aberrations and minimizing stray-light effects from reflections off surfaces of refractive elements. An important application is in the area of optical tweezers. At the same time, the axially non-symmetric OAPs introduce large geometrical field-dependent aberrations, primarily astigmatism, making the imaging of extended objects challenging. 

Two identical OAPs, oriented such that the axes of their parent parabolas coincide, form an optical relay, with the second OAP correcting the field-dependent astigmatic aberrations of the first~\cite{bruckner2010}. This design was first reported in~\cite{malone2006} for thermal imaging and adapted in~\cite{hejduk2019} as a microscope. The optical relay does not cancel distortion, in fact adding the distortion from the two mirrors~\cite{bruckner2010}. However, this does not affect the spatial resolution, which is limited by the aberrated point spread function size, and can be removed by an appropriate calibration and a mathematical transformation in software during post-processing. 

Previous works~\cite{malone2006} noted that the magnitude of the distortion is proportional to the focal length of the OAPs, and the OAPs' diameters can be chosen~\cite{hejduk2019} so that there would be no significant distortion in the desired field of view. In the case considered here, the optical arrangement is set due to experimental constraints, so microscopic distortion needs to be mapped and removed in software. To the best of our knowledge, this has not previously been applied to this optical design with micrometer-scale resolution and near the diffraction limit. The correction is non-trivial because the higher order distortion is severe and radially non-uniform, with limited calibration data available; so overfitting is a major concern. 

In this paper we perform distortion correction for the OAP relay microscope. We obtain a simple, readily-adaptable distortion model through model selection based on simulation data and test the model on experimental data from two distinct setups. 

\section{Experimental constraints}

The ability to provide the best possible Gaussian mode, with a diffraction-limited spot enabling high spatial resolution, is particularly important in an experiment using optical tweezers to support a dielectric microsphere (MS), whose position readout results in an exceedingly sensitive force sensor used to investigate new types of interactions at micrometer distance~\cite{venugopalan2024, blakemore2021, kawasaki2020}. For these measurements, a density-patterned "attractor" is scanned behind a stationary electrostatic shield $\sim\qty{3}{\micro\meter}$ from the surface of MS. Light scattered by the MS in the forward direction is recollimated and used to measure its position with a radial displacement noise of the order of $\sim\mathrm{pm/\sqrt{Hz}}$~\cite{venugopalan2024a}. 

Two $90^{\circ}$ off-axis parabolic mirrors (OAP~1 and OAP~2) in vacuum, with diameters and focal lengths of $\qty{50.8}{\milli \meter}$, are used to focus and recollimate the $\qty{1064}{\nano \meter}$ trapping beam of the optical tweezers ~\cite{kawasaki2020}.  OAPs are chosen to minimize halos that may scatter off the nearby attractor and shield and produce backgrounds. Good quality metrology is required to align attractor and shield relatively to each other, and to the position of the MS, with $\sim \qty{1}{\micro\meter}$ accuracy in the focal plane of the OAPs. While a commercial microscope objective is used in one view orthogonal to the trapping beam, the view along the trapping beam is particularly important for alignment and has to be obtained through OAP~2. This, uncorrected, introduces unacceptable aberrations over the required field of view of $\sim \qty{500}{\micro \meter}\times \sim\qty{500}{\micro \meter}$. For aberration correction, OAP~3 is added outside the chamber; the diameter of the OAPs and the path length are set by the existing optical tweezers system, so that distortion has to be corrected in software. 

\section{Off-axis imaging optics setups}

\begin{figure}[h]\centering
\includegraphics[width=\linewidth]{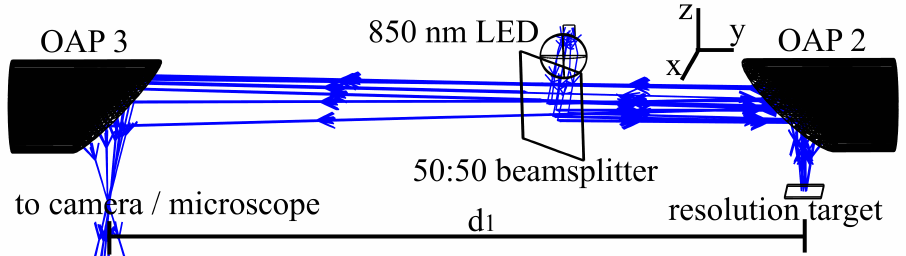}
\caption{Layout of test setups produced using Zemax OpticStudio. Distance between the OAPs ($d_1$) is $\sim \qty{86}{\centi \meter}$ for setup 1 and $\sim \qty{66}{\centi \meter}$ for setup 2. Folding mirrors, used for alignment and for orienting OAP~3 under physical constraints, are not depicted. Components are not drawn to scale. OAP~1 is located $\qty{101.6}{\milli \meter}$ below OAP~2 in the optical tweezers system. }
    \label{fig:layout} 
\end{figure}

Two test setups have been built with the layout in Figure \ref{fig:layout} to test the OAP relay microscope separately from the optical tweezers system. Setup 1 (setup 2) has the distance $d_1$ between OAP~2 and OAP~3 set to $\sim \qty{86}{\centi \meter}$ ($\sim \qty{66}{\centi \meter}$). Having data from the two distinct setups is beneficial as they are unlikely to share alignment errors. Light from a $\qty{850}{\nano \meter}$ LED reflects off OAP~2 and front-illuminates a 1951 USAF resolution target placed at its focal plane. The reflected light travels through the relay, where OAP~2 introduces and OAP~3 balances out the field-dependent astigmatic aberration. A camera is placed at the focal plane of OAP~3 and its image shown in Figure~\ref{fig:distorted}(a). Resolution across the field deteriorates rapidly with slight alignment errors in this setup \cite{hejduk2019}, so the minimization of such errors across the $\sim \qty{1}{\milli \meter} \times \qty{1}{\milli \meter}$ field serves as a check that the system is close to optimal alignment.

Assuming an aperture stop diameter of $\qty{10}{\milli \meter}$, Zemax simulations give an image space NA of $0.193$, so the diffraction-limited resolution of the system under the Sparrow criterion \cite{kubalová2021} is $\qty{2.07}{\micro\meter}$. However, since the relay gives no magnification \cite{malone2006}, the resolution is limited by the $\qty{4.8}{\micro \meter} \times \qty{4.8}{\micro \meter}$ pixel size. A $5 \times$ microscope, built from a 20x objective and a $f = \qty{50}{mm}$ aspheric lens, can be added between OAP~3 and the camera to address this, while reducing the field of view to $\sim \qty{200}{\micro \meter} \times \qty{200}{\micro \meter}$. The resultant image (Figure~\ref{fig:distorted}(b)) has a resolution of $\qty{2.19}{\micro\meter}$, similar to what is reported in~\cite{hejduk2019}. 

\section{Distortion correction}
\subsection{Radially non-uniform distortion}

\begin{figure*}[tb]
    \centering
    \begin{subfigure}[t]{0.495\textwidth}
    \includegraphics[width=\linewidth]{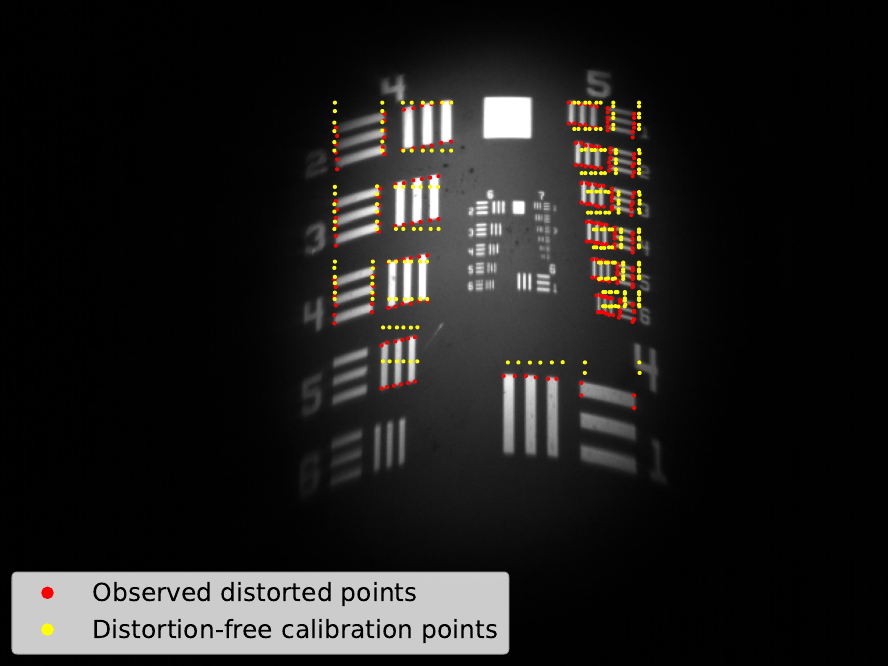}    
    \subcaption*{\makebox[\linewidth]{(a)}}
    \end{subfigure}
    \begin{subfigure}[t]{0.495\textwidth}
    \includegraphics[width=\linewidth]{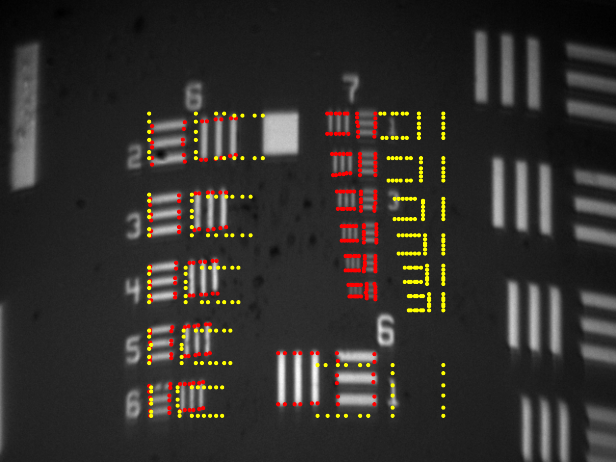}    
    \subcaption*{\makebox[\linewidth]{(b)}}
    \end{subfigure}
    
    \caption{(a) Image (flipped vertically) of the resolution target after the relay in setup 1, with groups 4 and 5 in the field of view. The camera has $640 \times 480$ pixels and a pixel size of $\qty{4.8}{\micro \meter} \times \qty{4.8}{\micro \meter}$. The two sets of points used for the correction, normalized so that they are aligned along the top and the left, 
    are labeled in red and yellow. (b) Similarly labeled image of the same target after the $5 \times$ microscope. Only the smaller groups 6 and 7, at the center in (a), are in the field of view. A scale bar for both images is provided in the distortion-corrected Figure~\ref{fig:undistorted}.}    
    \label{fig:distorted}
\end{figure*}

Zemax simulations of the relay (Figure~\ref{fig:simulations}) show that significant distortion is expected even if the system is optimally aligned, due to the long path length and the large field of view. Over a field of $\qty{1}{\milli \meter} \times \qty{1}{\milli \meter}$ in object space (approximately the extent of groups 4 and 5), the maximum distance between a distorted point and the corresponding paraxial point is $\qty{193}{\micro \meter}$, or $\qty{40.2}{\text{pixels}}$. The average distance is $\qty{58.3}{\micro \meter}$, or $\qty{12.1}{\text{pixels}}$, significantly larger than the 3 to 4 pixel perturbation assumed by \cite{weng1992}.

Simulations also show more distortion in $y$, which is along the off-axis portion of the OAP, than in $x$. A fit with a second-order radial model and a free distortion center leaves a large residual, with an average error of $\qty{33.6}{\micro \meter}$ ($\qty{6.99}{\text{pixels}}$). Radially non-uniform distortion has been reported and corrected in off-axis systems including Schwarzschild two-mirror hyperspectral imagers \cite{yang2021} and wide angle cameras for astronomy \cite{deppo2015}. Such distortion limits the correction methods available, since many of them, including some plumb line \cite{devernay2001} and some parameter-free methods \cite{hartley2007}, account for radially uniform distortion only \cite{tang2017}. In addition, many methods for detecting distortion centers assume radially uniform distortion \cite{tang2017} \cite{hartley2007}. 

\subsection{Model selection based on Zemax data}

Distortion in the OAP relay microscope can be corrected by selecting a distortion model with radially non-uniform components, calculating its coefficients based on a calibration image, and applying the transformation to new images. 

Since the resolution target cannot be installed in the optical tweezers system, only a few features are available in the focal plane for calibration. Hence, it is important that the distortion correction can be calculated with a reduced parameter set to avoid overfitting and allow good performance on new images without additional calibration. A model selection step is performed using simulated data to eliminate some parameters from the original distortion model. Those parameters will be set to 0 for subsequent corrections. The bias-corrected Akaike Information Criterion (AICc) \cite{spiess2010} is used to quantify the goodness of the nonlinear fit during this selection step. 

The Brown-Conrady model is chosen as the original model because it is commonly used and implemented in the open-source software OpenCV \cite{opencv}. It considers radial, decentering, and thin-prism distortion, interpreted physically as accounting for imperfect lens curvature, decentering of optical components, and tilt of lens with respect to the sensor, respectively \cite{weng1992} \cite{wang2008}.

\begin{figure*}[t]
    \centering
    \begin{subfigure}[t]{0.495\textwidth}
     \includegraphics[width=0.9\linewidth]{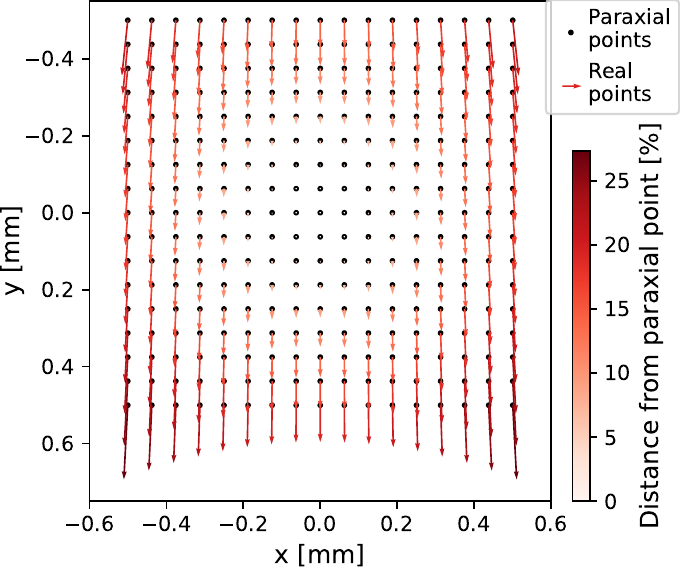}   
        \centering
    \subcaption*{\makebox[\linewidth]{(a)}}
    \end{subfigure} 
    \begin{subfigure}[t]{0.495\textwidth}
    
    \includegraphics[width=\linewidth]{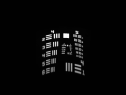}
    \centering
    \subcaption*{\makebox[\linewidth]{(b)}}
    \end{subfigure}
    \caption{Zemax simulations of setup 1 after the relay (both flipped vertically). (a) A $17 \times 17$ grid of paraxial image points are plotted using the Grid Distortion function, with quivers pointing to the corresponding real points. The color represents the distance between the two, normalized by field radius \cite{zemax2024}. The goal of the distortion correction is to undistort the real points so that they overlap the paraxial points. 
    (b) A digital image of the resolution target from \cite{thorlabs} is propagated  using the Image Simulation function and projected onto a $640 \times 480$ pixel grid to be compared to Figure \ref{fig:distorted}(a). 
 }
    \label{fig:simulations}
\end{figure*}

Let $\{(x_u, y_u)\}$ be the ideal distortion-free image coordinates and $\{(x_d, y_d)\}$ be the observed distorted coordinates. We then introduce $\bar{x}_u = m(x_u + x_\text{offset})$ and 
$\bar{y}_u = m(y_u + y_\text{offset})$ to account for magnification and relative position shift between the calibration image and the observed image, which cannot be calculated directly without knowing the distortion center. Similar to the parameterization in \cite{tang2017}, this allows the distortion center to be varied during the fit. We introduce $x_d = \bar{x}_d
+ c_x$ and 
$y_d = \bar{y}_d + c_y$ to enable a shift back to the original position in undistorted space. Defining $\bar{r}_u = \sqrt{\bar{x}_u^2 + \bar{y}_u^2}$, the Brown-Conrady model gives 
\begin{equation}\label{eqn:bc-model}
\begin{split}
\bar{x}_d &= \bar{x}_u + k_1 \bar{x}_u \bar{r}_u^2 + k_2 \bar{x}_u \bar{r}_u^4 \\&\quad + p_1(\bar{r}_u^2 + 2\bar{x}_u^2) + 2p_2 \bar{x}_u \bar{y}_u + s_1 \bar{r}_u^2 + s_2 \bar{r}_u^4, \\
\bar{y}_d &= \bar{y}_u + k_1 \bar{y}_u \bar{r}_u^2 + k_2 \bar{y}_u \bar{r}_u^4 \\&\quad+ p_2(\bar{r}_u^2 + 2\bar{y}_u^2) + 2p_1 \bar{x}_u \bar{y}_u + s_3 \bar{r}_u^2 + s_4 \bar{r}_u^4.
\end{split}
\end{equation}

The 289 pairs of paraxial and real points from the Zemax simulation shown in Figure~\ref{fig:simulations}(a) are used as data points for model selection. Denser points have been tried with no effect on the final selected model. The simulated points have no noise, a known distortion center at $(0, 0)$, and a magnification of 1. To mimic the experimental situation, where the distortion center is not known, the points are re-normalized so that their coordinates have values between $\pm 0.5$ along the $y$ direction while maintaining the aspect ratio. Because of the small initial distortion in $x$ direction, the distortion centers of the two sets of points are still close after re-normalization, so the parameter $x_\text{offset}$ has to be added to the final model. 

The original model in Equation~\ref{eqn:bc-model} has 13 free parameters ($k_1, k_2, p_1, p_2, s_1, s_2, s_3, s_4, x_\text{offset}, x_\text{offset}, c_x, c_y, m$). The magnification cannot be zero, but each of the other 12 parameters can be included or excluded in a possible model, resulting in $2^{12} = 4096$ models if stepwise model selection \cite{whittingham2006} is avoided. 
For each model, the initial value of every parameter is set to 0, except for the magnification, which is set to 1. Distortion coefficients are calculated with a nonlinear fit and used to numerically undo the distortion. The AICc of the distorted-corrected points compared to the corresponding calibration points are calculated, and the model with the lowest AICc is selected. This model has finite values for the parameters $k_1, p_1, s_3, y_\text{offset}, c_x, c_y, m$, while the rest are set to 0. Adding in the $x_\text{offset}$ mentioned above, the simplified model is 
\begin{equation}\label{eqn:selected-model}
\begin{split}
\bar{x}_d &= \bar{x}_u + k_1 \bar{x}_u \bar{r}_u^2 + p_1(\bar{r}_u^2 + 2\bar{x}_u^2) \\
\bar{y}_d &= \bar{y}_u + k_1 \bar{y}_u \bar{r}_u^2 + 2p_1 \bar{x}_u \bar{y}_u + s_3 \bar{r}_u^2.
\end{split}
\end{equation} 

\subsection{Distortion correction on experimental data}
The model selected based on Zemax data should apply to all points from both setups, with or without the $5 \times$ microscope, since a well-corrected microscope adds negligible radial distortion.

To verify this, the corners of line pairs on the resolution target are labeled for images taken on each test setup, yielding sets of observed distorted points. The corresponding sets of calibration points are found by imaging the same resolution target under a commercial optical microscope. Since the calibration points are from a different imaging system, their coordinates are normalized so that their range in the $y$ direction is comparable to that of the distorted points, and the origins of the two sets of points are overlapped. The normalized points are shown in Figure~\ref{fig:distorted}. Note that because the images are severely distorted and near the diffraction limit, automatic corner detection is difficult and the corners are labeled manually using \cite{dutta2019}, which is viable because the optical tweezers system is fixed and rarely needs to be calibrated. This yields 238 pairs of points for setup 1 and 288 pairs of points for setup 1 after the $5 \times$ microscope and setup 2. 

\begin{figure*}[t]
    \centering
    \begin{subfigure}[t]{0.495\textwidth}
    \includegraphics[width=\linewidth]{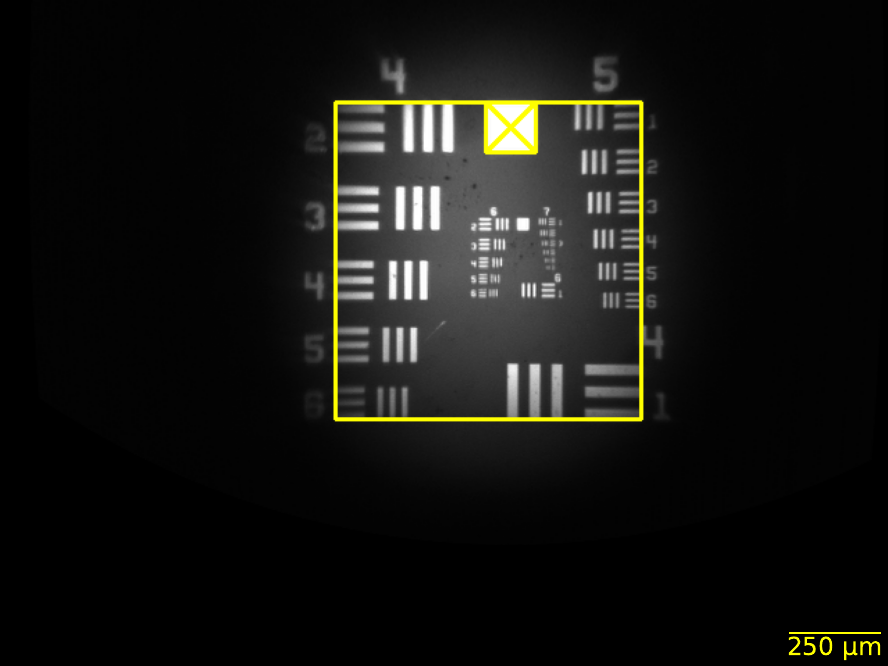}    
    \subcaption*{\makebox[\linewidth]{(a)}}
    \end{subfigure}
    \begin{subfigure}[t]{0.495\textwidth}
    \includegraphics[width=\linewidth]{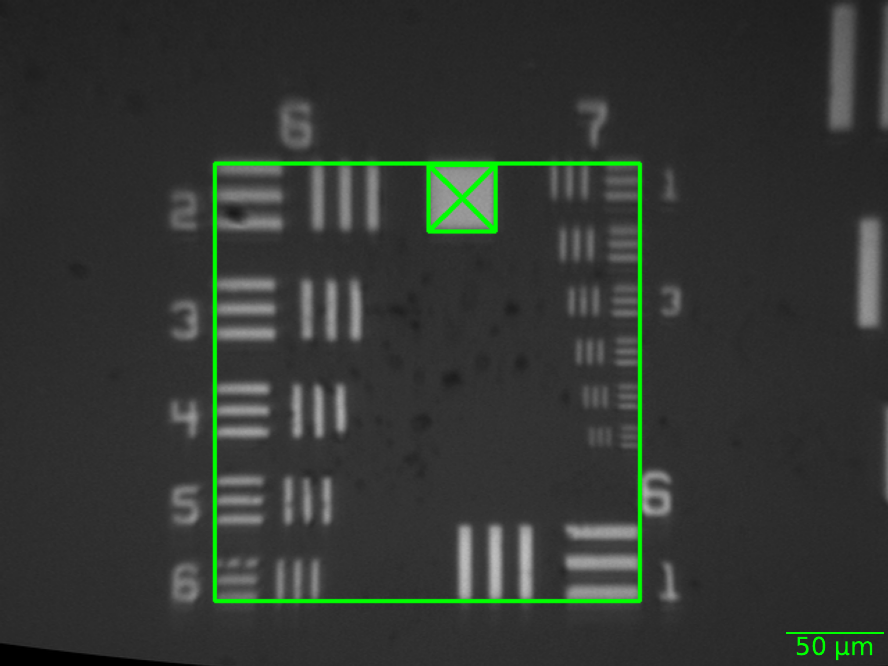}    
    \subcaption*{\makebox[\linewidth]{(b)}}
    \end{subfigure}

    \caption{Distortion-corrected images. (a) Image after the relay (corrected Figure~\ref{fig:distorted}(a)). The yellow rectangles are overlaid on the edge of the groups to check that the distortion-corrected edges are straight. The squares with cross bars are overlaid on squares on the resolution target to check that the aspect ratio is as expected. (b) Image after the $5 \times$ microscope (corrected Figure~\ref{fig:distorted}(b)), with the green rectangles and the squares similarly overlaid.}
    
    \label{fig:undistorted}
\end{figure*}

The distortion coefficients for each setup are then calculated, with the initial guesses adjusted to be of the correct magnitude in pixels ($k_1=10^{-6}, p_1=10^{-4}, s_3=10^{-4}, x_\text{offset}=-320, y_\text{offset}=-240, c_x=320, c_y = 240, m=1$) since coordinates are now on the order of $10^2$ pixels. After obtaining the coefficients, we use OpenCV to perform the distortion correction, producing the corrected images shown in Figure~\ref{fig:undistorted}. Future images can be simply corrected by applying the same transformation.

\begin{table}[htbp]
\centering
\caption{\bf RMS residual distortion, after corrections, for the various optical setups discussed }
\resizebox{\linewidth}{!}{
\begin{tabular}{b{0.3\linewidth}C{0.16\linewidth}C{0.15\linewidth}C{0.3\linewidth}}
\hline
setup & max error [px] & average error [px] & average error in object space [$\unit{\micro\meter}]$\\
\hline
setup 1, after relay & 2.05 & 0.93 & 3.60
 \\
setup 2, after relay & 3.69 & 1.24 & 4.11\\

setup 1, after $5 \times$ microscope & 4.49 & 1.84 & 1.33 \\

\end{tabular}}
\label{tab: results}
\end{table}

To quantify the quality of the corrections, the RMS residual error of the points is reported in Table~\ref{tab: results}. Following \cite{tang2017}, given $M$ pairs of distorted and distortion-corrected points $\{(x_{d_i}, y_{d_i})\}, \{(x_{u_i}, y_{u_i})\}$ and letting $\{(f_x(x_{d_i}, y_{d_i}), f_y(x_{d_i}, y_{d_i}))\}$ represent the corrected points, we define total residual error as $\sqrt{\sum_M{((f_x(x_{d_i}, y_{d_i}) - x_{u_i})^2 + (f_y(x_{d_i}, y_{d_i}) - y_{u_i})^2})}$, and convert the error to object space based on the known sizes of features on the target. With the $5 \times$ microscope, the residual error in object space decreases significantly with the tradeoff of a smaller field of view.

 One contribution to the residual error derives from diffraction at the edges of each line pair in the image, which adds uncertainty to the observed locations of the distorted points. This can be quantified in the distortion-corrected image by assuming a Gaussian PSF and fitting the edge spread functions \cite{smith2008} to the error function. In setup 1, the error function has a maximum $\sigma$ of $\qty{1.57}{\text{pixels}}$ and an average of $\qty{0.76}{\text{pixels}}$. After the $5 \times$ microscope, it has a maximum $\sigma$ of $\qty{3.73}{\text{pixels}}$ and an average of $\qty{1.55}{\text{pixels}}$, which likely contributes significantly to the residual error. 
 
 The performance of the reduced-parameter model is also tested against the full model described in Equation~\ref{eqn:bc-model} through simulations. Because the goal is to reduce numerical risk such that the model successfully converges with fewer data points, as is necessary for the optical tweezers system, $90\%$ of available experimental data points are randomly selected in each trial for the fit. In 10,000 trials, the fit with the reduced-parameter model does not converge 0 times (9 times) for setup 1 (setup 1 with the $5 \times$ microscope). The full model with 13 parameters does not converge 4130 times (1186 times), presumably because of overfitting. For setup 2, both models perform well.

\section{Conclusion}
In this paper we correct for the higher-order distortion of an OAP relay microscope with $\qty{850}{\nano \meter}$ illumination. To prevent overfitting, we select a simple distortion model using AICc based on simulated data. Depending on whether an additional $5 \times$ microscope is used, corrected images from two test setups achieved a field of view of $\sim 1\times \qty{1}{\milli \meter}$ with a mean residual error of $\qty{3.60}{\micro\meter}$ or a field of view of
$\sim 200\times \qty{200}{\micro \meter}$ with a mean residual error of $\qty{1.33}{\micro\meter}$, respectively. Part of these distortion errors are likely due to diffraction artifacts in the features used for the correction, since the system is near diffraction-limited. 

\begin{backmatter}
\bmsection{Funding}
NSF Grant No. 2406999, ONR Grant No. CON-80004489, and the Heising-Simons Foundation. 

\bmsection{Acknowledgment} We thank Professor Daewook Kim of the University of Arizona for helpful discussions on aberration and distortion from OAPs and for a careful read of the manuscript. We also acknowledge discussions with Clarke Hardy, Chengjie Jia, Kenneth Kohn, Lorenzo Magrini, Gautam Venugopalan, and Zhengruilong Wang. M.L. wishes to acknowledge her support by the Stanford Summer Undergraduate Research Program.

\bmsection{Disclosures} The authors declare no conflicts of interest.

\bmsection{Data availability} Data underlying the results presented in this paper are available in the \href{
https://github.com/stanfordbeads/oap_relay_microscope_distortion}{Github repository}.

\end{backmatter}

\bibliography{references}
\bibliographyfullrefs{references}

\end{document}